# Discovering adoption barriers of Clinical Decision Support Systems in primary health care sector

*Soliman S M Aljarboa*

Qassim University, Buraydah 52571, Saudi Arabia

*Shah J Miah*

Newcastle Business School, University of Newcastle, NSW, Australia

**Abstract**

Adopting a good health information system (HIS) is essential for providing high-quality healthcare. With rapid advances of technology in the healthcare industry recent years, healthcare providers seek effective options to deal with numerous diseases and growing number of patients, adopting advanced HIS such as for clinical decision support. While the clinical decision support systems (CDSS) can help medical personnel make better decisions, they may bring negative result due to a lack of understanding of the elements that influence GPs' adoption of CDSS. This paper focuses on discovering obstacles that may contribute to the problems surrounding the CDSS adoption. Thirty general practitioners were interviewed from different primary health centres in Saudi Arabia in order to determine the challenges and obstacles in the sector. While the outcome confirms that there are obstacles that affect the aspects, such as time risk, quality of the system used, slow Internet speed, user interface, lack of training, high costs, patient satisfaction, multiple systems used, technical support, computer skills, lack of flexibility, system update, professional skills and knowledge, computer efficiency and quality and accuracy of data.

**Please cite as:** Aljarboa, SSM., Miah, S.J. (2022). Discovering barriers of a clinical decision support system adoption in primary health care sector: a qualitative study, Health Information Management Journal, (under review).

1. Introduction

An advanced and supportive medical systems to assist doctors in making decisions has been a vital need in healthcare industries. One of the many reasons for that is large number of patients with a wider range of healthcare conditions. The clinical decision support systems

(CDSS) supports this process because they provide alerts and recommendations to end-users, helping them to make the right decisions. However, failure to take into account of the factors that undermine the adoption of CDSS seriously may impact to the weakness or failure of any support strategy. This motivates us to investigate the aspect of CDSS adoption specifically in a developing country context. We found that an in-depth interview technique to collect the qualitative insights would be suitable in the target context of our study. For the semi-structured interviews, we conduct the study with GPs currently working in Saudi Arabia's primary healthcare centres. Existing studies are limited in this that investigated and studied CDSS adoption by healthcare professionals in Saudi Arabia (Alqahtani, Alshahri, Almaleh, & Nadeem, 2016; Alsulame, Khalifa, & Househ, 2016).

Primary healthcare has been rapidly transformed in developing countries. For instance, Al Asmri, Almalki, Fitzgerald, and Clark (2020) noted that the primary healthcare system in Saudi Arabia are in a process of reforming its processes for delivering effective care services. Their study suggests that challenges in the healthcare system are to be addressed with appropriate technological solutions, improving practices such as equitable access to the services, ensuring quality and safety of services, as well as modernising current referral systems (Al Asmri et al., 2020, p. 473). In a similar study conducted at the general clinics of primary healthcare centres, it is indicated that patients were not satisfied enough (53% dissatisfaction) with their healthcare consultations due to lack of transparency (Tabekhan, Alkhaldi, & Alghamdi, 2018). This implies that physicians should improve their practices for better medical consultation. For this reason, it may be important to adopt CDSS, which can contribute significantly both to enhancing patients' satisfaction and supporting physicians in providing more accurate and appropriate medical consultations (Shaker & Samir, 2014). In Riyadh, another study evaluated patient opinions regarding the existing quality of primary healthcare; this study revealed that most patients did not choose primary care centres for care and treatment, which may result due to lack of confidence (Alzaied & Alshammari, 2016).

CDSS has many capabilities and advantages but these have not been exploited due to poor design. This is generally because designers do not take into account the factors that influence the adoption of CDSS (Khairat, Marc, Crosby, & Al Sanousi, 2018). Studying and investigating the adoption of CDSS is of paramount tasks and would be contributing significantly, to the industry practitioners and academic researchers through new understanding on different end-user perceptions and insights above innovative use of CDSS in healthcare sector (Wijnhoven & Koerkamp, 2019). The next section presents a detailed background of the study reported in the paper.

## 2. Literature review

A CDSS is a computerised information management system developed to help healthcare professionals to make effective decisions concerning patient care and medication (Keine, Zelek, Walker, & Sabbagh, 2019). CDSS is used for medical prevention, diagnosis and treatment, enabling better clinical decisions and providing better healthcare services. Many research have been introduced in this domain. For example, Holstiege, Mathes, and Pieper (2015) evaluated the influence of using CDSS in enhancing antibiotic prescriptions in primary patient care. Their results suggested that CDSS provides a significant positive outcome regarding antibiotic prescriptions. The authors called for more research on CDSS design and assessing its ability to enhance acceptance and adoption by the relevant users.

The challenges in adoption IT systems are greater in developing than developed countries (Esmaeilzadeh, Sambasivan, Kumar, & Nezakati, 2015). This is attributable to a lack of human and medical resources, computer illiteracy, the poor design of human-technology interfaces and a lack of training and experience in developing countries (Zakane et al., 2014). Adoption of CDSS in rural healthcare in developing countries faces certain challenges, for example, weak electrical power current, computer skills and knowledge necessary for the correct use of the medical system (Zakane et al., 2014). Other significant obstacles also include many alerts, the cost of introducing CDSS into healthcare organisations, and the implications for privacy, confidentiality and information management efficiency (Gullapalli, Brungi, & Gopichand, 2015). The results indicate that challenges listed are mainly: financial and organisational aspects. The authors highlight the importance of seeking experienced help from external sources to implement and examine HIS, along with the importance of working on compatible workflows to ensure the healthcare application success. They add that it is necessary to define a timeframe for implementation and allocate appropriate budgets for various stages of system implementation.

The lack of interaction between end users and the system can be an issue that affects CDSS adoption. With reference to physicians and CDSS regarding decision-making, this issue is a major reason for its poor implementation (Miller et al., 2015). The difficulties of using CDSS can result in unsuccessful outcomes (Jung et al., 2020). For example, Laka, Milazzo, and Merlin (2021) indicated that participants (doctors in a hospital and primary care centres in Australia) may play important role if CDSS system is flexible and easy to use. If the CDSS provides insufficient information, the outcomes may be negative (Castillo & Kelemen, 2013). Wright et al. (2016) indicated the system may have some technical problems that are difficult for users to notice and may persist for a longer time, for several reasons including software upgrades, modifications to codes and fields, and unintentional deactivation. On the other hand, Johnson et al. (2015) concluded that data entry may be difficult in CDSS that can make structural and practical obstacles to effective use the system.

Aforementioned discussions indicate that CDSS adoption can be a complicated task which has been an important issue in healthcare sector. The existing studies have attempted to address the issues in different ways for revealing the assistive insights, but a holistic approach is of paramount to apply for bringing quality outcome. The new outcome through comprehensive research can only provide effective support for researchers and practitioners in this sector. We therefore believe that our attempt can be imperative in the field of research. Following section presents the methodological explanation of the study reported in this paper.

## 3. Research methodology

This study aims to discover the obstacles and challenges that affect the adoption of CDSS. We adopt a qualitative research method, because the qualitative methods are suitable for an in-depth comprehension of certain phenomena and offer richer information, using statements from target participants (Merriam, 2014). When qualitative data are analysed, it usually leads to a deeper understanding of the problem, therefore enables the researcher to obtain more abundant and powerful information in comparison to quantitative data methods (Anderson, 2010). The semi-structured interview offers freedom to the participant through open questions, imparting confidence to the interviewer. This contributes to obtaining more reliable data through in-depth discussion and debate on certain themes or issues (McIntosh & Morse, 2015). According to Yin (2014), the semi-structured interview must have enough participants to reveal the richness of people's perceptions and obtain comprehensive comparisons.

The use of qualitative research methods, specifically semi-structured interviews with 30 GPs, was deemed the most appropriate method for collecting data owing to the need to explore the perceptions and expectations of the sample population regarding the obstacles they experience concerning the adoption of CDSS. This approach will greatly contribute to understanding the obstacles through in-depth interviews with GPs, which quantitative research cannot identify. Following this research method will contribute to providing rich data from GPs as an end user to help healthcare system policymakers and system providers avoid the problems that hinder CDSS adoption.

### 3.1 Data collection and analysis

Thematic analysis has been employed to analyse data obtained from interviews with GPs. Objective analysis helps in discovering topics and helps in organising and arranging data (Maguire & Delahunt, 2017). Moreover, Nowell, Norris, White, and Moules (2017) suggested that thematic analysis contributes to extracting unexpected insights and ideas from rich data and helps in summarising the rich data. Thematic analysis has become widely employed in HIS studies (De Leeuw, Woltjer, & Kool, 2020; Stapley et al., 2018).

Thematic analysis contributes to the formation of theories through a set of procedures that generate facts and factors (Guest, MacQueen, & Namey, 2012). The data were then mainly analysed using NVivo software, following the six phases by Braun and Clarke (2006) to analyse the qualitative data: (1) familiarisation with the data; (2) generating initial codes; (3) searching for themes; (4) reviewing themes; (5) defining and naming themes; and (6) producing the report.

The first step is familiarising data: the transcripts of the interviews were read, and the main ideas underlined, and the records were listened to more than once. Notes were taken to recognise the existing data and to better understand the interviews. At the second step (generating initial codes), a list of different words was encoded from the transcripts, which categorised the data into themes. NVivo 12 software served to formulate the different codes. The third step is searching for themes and this involves intensive work to classify and link the themes with the coding undertaken in the previous step.

### 4.1 Sampling

A total of 30 GPs have participated in this study. The GPs are currently working in healthcare centres in the two major cities of Riyadh and Qassim, under the supervision of the Ministry of Health in Saudi Arabia. This paper employed both purposive and snowball sampling methods to choose those taking part. The purposive sampling methodology is designed specifically for collecting in-depth experiences in qualitative research (Etikan, Musa & Alkassim, 2016). The participants are predefined, with the aim of identifying people who can offer information during the interviews. The main features for analysis include personal characteristics and participants' depth of understanding of the central themes (Ritchie, Lewis, McNaughton Nicholls, & Ormston, 2014). Sharma (2017) define purposive sampling as a technique used mainly in qualitative studies to select participants by evaluating their specific purpose according to what the research questions ask. In addition, specific participants are drawn from the target population, with the research study objective being a key point of consideration.

4. **Results and discussion**

As mentioned earlier we have interviewed 30 general practitioners under the ethical approval[1] by ethical committee in an Australian university. The outcome contributes to obtaining important views that helped discover the obstacles that may better guide the adoption of CDSS. After conducting the data analysis, 14 obstacles were identified that health system providers and developers must take into account in order to implement CDSS successfully. In this section, we present and discuss each factor that was discovered with statements given by the GPs. In an overall picture, all participants noted the importance and positive impact of the CDSS and its role in providing better healthcare. They believed that the positive impact of the CDSS outweigh its negatives. However, they pointed out that it is important to address the negatives and challenges in order to successfully adopt the system and take advantage of its features.

**4.1 Time Risk**

Time risk is one of the main factors that may hinder the adoption of the CDSS. GPs indicated that the time spent using CDSS during the diagnosis and interviewing patient involved a lot of data entry (necessary and unnecessary) that may prolong patient diagnosis and sitting with him or her, and this simply adds to the waiting time for other patients. Some GPs were concerned that the use of such advanced medical programs may increase time spent because of using more than one health application, or that they may need to fill in fields to record a diagnosis and obtain support for making a medical decision. As stated by one participant:

*"The use of many medical programs will create a problem, unfortunately. We are currently using various medical applications such as the electronic health record program and the drug programs and if yet another program is used that helps with a diagnosis, together these mean more work for the medical practice and this will lead to more time being spent"* (GP9)

Furthermore some GPs felt that writing might also cause more time to be spent on the CDSS:

> *"I think sometimes such programs take a lot of time. I mean, for example, in order to benefit from applications such as this, you should add all the requested data like diagnosis, medical history, and the tests I have done. Therefore, because you're adding complete data to each patient, this process takes a lot of time."* (GP10)

GP10 added that she would use CDSS if she discovered that it was beneficial to the patient, even if it is time-consuming:

---
[1] **Ethical Approval REF. No: HRE18-008**

> *"For me, the most important thing I have is the patients: if I know this will help them in diagnosis and healthcare, I will use it even if it takes a lot of time. I am sure I will use it."*

Arts et al. (2018) investigated obstacles concerning the use of CDSS in general practice. The study reported that the principal obstacle and challenge was lack of time during the consultation and patient diagnosis. In another study, Rieckert, Sommerauer, Krumeich, and Sönnichsen (2018) evaluated the PRIMA-eDS-tool, which is an electronic DSS for GPs to reduce inappropriate medications and provide medical advice regarding medications. The findings showed that GPs perceived the data entry in the system for each patient to be annoying and time-consuming, which could simply undermine people's adoption. Asan, Carayon, Beasley, and Montague (2015) investigated the factors related to the use of EHR and the results indicated that the time consumption factor was one of the main reasons of why physicians did not share the screen with the patient. There simply was not enough time for this and it could cause unnecessary distractions. For this reason the physicians indicated the need to design CDSS effectively and provide training to improve communication with the patient.

**4.2 System quality**

The quality of the system in terms of interruption or suspension of the medical application is one of the obstacles that affect CDSS adoption. They also claimed that this problem would impede their work performance, which may have serious consequences for patients' health. They stressed the importance of designing an effective and high-quality CDSS. Participants support this position:

> *"We need a strong infrastructure so not causes a problem in suspension or breakdown of CDSS due to the crowding out patients… The problem may arise because a large number of users on a web-based system, especially at peak time, the system is at its lowest level, meaning that there is a problem with either the bandwidth or the servers. Then it resumes working after the rush hour."* (GP7)

Another participant stated that:

> *"One of the challenges that may affect the workflow is the system stoppage, as is the case in the current system at times. This causes a long wait for the patient and postpones medical prescriptions until the system works again."* (GP4)

Deborah (2014) identified several major challenges of the implementation of EMR. The author stated that the paucity of infrastructure was considered as the main reason for suspension and disconnection of the HIS, and thus hindered medical work. This could result in a strained relationship between the patient and physician. Bouamrane and Mair (2013) studied GPs' perspectives of using HIS in primary healthcare centres in Scotland. Those participants, however, stated that the failures of the system had happened only a few times or seldom. In another qualitative study, Bouamrane and Mair (2014) investigated the perspectives of GPs regarding the role and effectiveness of e-referrals. Results showed that the benefits of using e-referrals in primary care centres significantly outweighed the negative aspects. Nevertheless, many GPs in that study complained about their frustrations caused by slow connections and breakdowns in the system, which caused delays or failures to complete the referrals electronically.

## 4.3 Speed Internet connection

The Internet connection is claimed to be one of the main requirements for providing better medical care in the modern world, however still neglected in many developing country context. As many medical applications require a faster Internet connection. Yet, in some circumstances the speed of the Internet can be slower, which may cause of delays in operating the system or difficulty in entering or extracting patient data:

*GP18 said "Internet speeds may be slow, which negatively affects the use of the current medical system, which in turn affects work performance."*

And *GP22 pointed "The Internet should not be slow, as this contributes to taking longer time in the medical work and leads to the inability to use the medical system properly."*

Zaman, Hossain, Ahammed, and Ahmed (2017) pointed that high-speed Internet has a positive influence on the success and implementation of HIS. The availability of easy access to the Internet can positively affect the user in finding health information (McCloud, Okechukwu, Sorensen, & Viswanath, 2016). On the other hand, slower Internet in medical health centres may make doctors frustrated and dissatisfied with the electronic system, and this may affect their performance of medical tasks (Ajuwon, 2015).

## 4.4 Complex user interfaces

An important aspect of accelerating end-user adoption is how simple and uncomplicated the user interface of a medical application is. Some doctors suggested that the complexity of the user interface through many requirements or the introduction of unnecessary data commands, all contribute to preventing the wider adoption of the CDSS:

*"The developers should make the interface of the software smooth and simple, that is, they are neither complicated nor duplicated, for example, what you require in the beginning is not what you will ask again eventually. It's also better to avoid repetition." (GP24)*

A weak design of the user interface, which ideally should increase access to the relevant information that contributes to making decisions, may slow down the physicians' work. It is important that HIS interfaces increase end-user confidence, are easy to use and not be time-consuming (Tsopra et al., 2019). Aljarullah, Crowder, and Wills (2017) studied the factors that influence primary care physician acceptance of HER. They noted that perceived ease of use is a significant influence on physicians' acceptance of HER. Koskela, Sandström, Mäkinen, and Liira (2016) noted that it is important in implementation of CDSS that the user interface design is easy to use, and that dealing with it does not take time. Testing the CDSS before its final adoption makes it possible to identify and explore problems that may appear in the user interface, which leads to avoiding harm and improving the functioning of the medical system (Madar, Ugon, Ivanković, & Tsopra, 2021).

### 4.5 Lack of training

The lack of training of GPs on how the medical system works is one of the reasons that discourage the adoption of the CDSS. Training is an important factor in facilitating and accelerating the use of the CDSS. The GPs indicated that the lack of training is a serious hindrance to the adoption of the CDSS. In order to ensure that the medical application is implemented correctly, healthcare providers must ensure the preparation of courses and adequate training for the end user.

*"Training is important as we need a lot of training to know how to optimally use the system." (GP15)*

*"Training is important for users of the system because there are usually some ambiguities and problems at the beginning of the use of the new medical system." (GP9)*

Several studies have discussed the importance of training for the successful implementation of HIS (Al Aswad, 2015; Thurston & Mulberger, 2015; Turan & Palvia, 2014). User training is important in improving computer self-efficacy and maximising IT use. The lack of training on using CDSS may cause the implementation of the system to fail and lead to no benefit from its assumed capabilities (Castillo & Kelemen, 2013). Hossain, Quaresma, and Rahman (2019) claimed that policies and plans should be developed to encourage and help healthcare practitioners to use the HIS in Bangladesh through proper training. Irfanahemad, Nandakumar, Ugargol, and Radhika (2018) evaluated the determinants that influence the use of telemedicine in a tertiary care paediatric centre. This implies that appropriate training and improved skills and capability for teleservices were necessary for healthcare employees.

### 4.6 Costs

The high cost of the HIS, especially high-quality systems, is one of the main factors that affects the adoption of the CDSS. High quality medical applications are expensive and require financial support or highly qualified and experienced software developers. One of the participants in this study pointed out that medical programs are expensive, and healthcare providers require financial support in order to adopt efficient and high-quality applications.

*"I think that the CDSS will be expensive because it needs an integrated team such as developers and consultants, especially when the program is new because we will start from scratch." (GP12)*

*"usually the cost of the systems will be high, but its echoes on health will be better, of course." (GP4)*

In relation to cost it should be noted that expenditure on CDSS development is undoubtedly high (Kosmisky, Everhart, & Griffiths, 2019). Dalaba et al. (2015) noted that, despite its high costs, CDSS is nevertheless effective and has valuable results. According to Alharbi, Atkins, and Stanier (2015), the complexity and high cost of e-health programs and low-level IT skills among practitioners are major barriers to their realisation. Alkraiji, Jackson, and Murray (2016) argued that switching costs of HIS is one of the main factors that influence the adoption of health data standards in Saudi healthcare organisations. Furthermore Zaman et al. (2018) examined three hospitals in Makkah to determine the most important obstacles facing HIS implementation. The study results showed that the one of the main issues was the high costs associated with IT.

### 4.7 Perceived patient satisfaction

One of the important factors that could affect the adoption of the CDSS, as indicated by some participants in the study, is the extent of patients' satisfaction and attitudes towards the system used. A patient's satisfaction are very important for assessing the quality of services and healthcare being provided. The GPs indicated there are some fears and concerns about the reliance on the CDSS to make decisions that may affect a patient's satisfaction and confidence in the physician. It is possible that eye contact with the patient will wane, and the GPs will focus more on the computer:

*"If the physician focuses much on the system, the focus will decrease in terms of dealing with the patient, the eye contact can affect a lot and by looking and interacting with the patient may provide more suitable diagnosis." (GP14)*

*"Some patients resent the doctor's use of the medical system! Because this takes time: for example, a patient comes and asks about the importance of using this system, and that it is the reason for wasting time, because the user takes approximately ten minutes to take his data. Although I try to shorten the*

*time and enter data after the patient is gone, but some of them did not accept the idea of using the computer and this affects the patient's confidence in medical care." (GP16)*

When adopting the HIS, it is important to take into account the extent of patients' satisfaction with the system used (Abdekhoda, Ahmadi, Gohari, & Noruzi, 2015). Amann, Blasimme, Vayena, Frey, and Madai (2020) pointed out that relying on artificial intelligence in medical diagnosis may have shape the relationship between doctor and patient in terms of trust.

### 4.8 Systems integration

A physician's reliance on more than one medical application alongside CDSS could lead to hesitation of CDSS adoption. Most GPs indicated they are currently using more than one medical application during the patient's visit, and this means that when medical corporate software is not integrated to CDSS, it may cause confusion. The presence of a sensibly integrated CDSS that includes more services and benefits can contribute to speeding up the time of use and faster and positive adoption by the GPs.

> *"I think integrating all programs into one system is very important because moving between one program and another wastes time, so integration will help to reduce the time."* (GP19)

> *"All programs must be integrated into one program, so merging helps to accept CDSS. I think that once I enter the patient identification number, I should find a place for X-rays, laparoscopy and old medicines. I am also supposed to find a place through which I can dispense new medicines, and read the old notes made by the physicians in the clinics the patient has visited."* (GP4).

The presence of many investment companies in implementing HIS has contributed to providing different medical applications that are relatively difficult to integrate because of such diversity in technical systems and their varying specifications (Payne et al. 2019). This leads to difficulty in sharing information and data between hospitals when software/hardware systems cannot 'talk' to each other. Al Aswad (2015) indicated that the laboratory system often uses different EMRs in many Saudi hospitals, where there is no integration of laboratory systems with EMR owing to unwanted or unplanned for complexity.

Study found that a combination of CDSS along with EMR provides better outcomes regarding care for diseases. Integrating CDSS with EMR contributes to the provision of directions, recommendations and alerts (Chatzakis, Vassilakis, Lionis & Germanakis, 2018). Integration of medical systems helps to make the system more flexible and thus reducing the time, cost and increasing accuracy (Kitsiou, Matopoulos, Vlachopoulou, & Manthou, 2009). Further, Bologva, Prokusheva, Krikunov, Zvartau and Kovalchuk (2016) highlighted that combining the systems positively affects the provision and access to data, higher quality communication, reduced medical errors and improved healthcare services.

### 4.9 Technical Support

The presence of highly qualified technical support available during working hours is a necessity to have successful adoption of the CDSS. The GP indicated that there is a fear that when there is a technical defect in the CDSS, there may be a delay in the arrival of appropriate technical support for the problem, especially since the technical support for the primary health centres is in places far from them:

> *"I think we must have existing technical support: IT technicians are supposed to be around or very close because or we may need them to modify any information or if there is any malfunction of the program or discontinue."* (GP24)

> *"...any system such as these systems is subject to any error or problem and could stop functioning at any time. So, it is important that there be technical support for these in case anything happened to these.*" (GP22)

Rahimi et al. (2018) added technical support factor to their study models to building a model that helps achieve a better understanding of HIS adoption. Grout et al. (2018) investigated CDSS tools that have contributed to supporting user acceptance. They mentioned that the active technical support teams are critical factor to achieving a successful long-term CDSS use. Furthermore, Lugtenberg et al. (2015) stated that the lack of technical support required, is considered to be as one of the principal perceived barriers to use the CDSS

### 4.10 User's computer skill

Personal computer skills are an important job for the end user to adopt and implement the CDSS quickly and easily. The physician who relies on the paper-based system in diagnosing the patient, especially if he uses it for a long time, may mean slow data entry or optimal use of the CDSS, which needs enough time to use it optimally and better.

*"The doctor should have a good skill in using computer, We are now in a period of time where the doctor must be familiar with using computers, I mean - he or she should be familiar with the extent that makes him deal with programs well."* (GP21)

*"The doctor needs to have good computer skills in order to be able to perform the system better."* (GP8)

*"It is important for the doctor to have some computer skills, especially basic knowledge, in order to perform the work in the medical application well."* (GP25)

The lack of computer skills for end users is one of the barriers that negatively affect the functioning of health systems (Alshami, Almutairi, & Househ, 2014). Zakane et al. (2014) asserted that the necessary and sufficient computer skills in order to use the system and take advantage of its operational capacities is a serious problem for CDSS in developing countries. Moreover, Alharbi et al. (2015) pointed out that doctors' lack of computer skills is one of the reasons for failing to achieve the required efficiency of using medical programs. It is important to increase the skills of health practitioners and other employees in computer skills, as this will contribute significantly to increasing EMR adoption (Hasanain, Vallmuur, & Clark, 2015).

### 4.11 Lack of flexibility

The flexibility of the systems used is an important feature that helps the physician to embrace the HIS faster. When the CDSS is flexible and helps GPs to make decisions and not force them to make certain choices in diagnosis, this contributes to the success of CDSS.

*"Sometimes the work system is inflexible when there are a lot of patients as it is faster and easier to prescribe medication on paper. As for the system, we already have to fill in the data, the fields and the details. I prefer a paper-based prescription when there are a lot of patients. For example, if you see 20 patients every 8 hours, you will have more flexibility." (GP26)*

*"The system should not be complicated, that is, I mean that filling the blanks in the medical program is not a lot and consumes time , but I want it to be fast and for the information to be available in a short time." (GP3)*

*"I think one of the annoying challenges is that the program limits the doctor's freedom of action at times and narrows the diagnosis, as it is necessary in some of such this software to adhere to certain codes and options." (GP20)*

It is important for the systems to be useful in supporting physicians' decisions so that such decisions do not jeopardise the autonomy of a patient (Sutton et al., 2020). Despite the benefits it provides to a CDSS, these systems are not considered as being flexible enough. In patient diagnosis, especially when there are several disease symptoms or rare diseases (Lodh, Sil, & Bhattacharya, 2017). Moreover, Holanda et al. (2012), this indicates that although most participants in the study believe that working on the EHR is better, paperwork is actually done faster than the EHR for documenting a medical task.

### 4.12 System updating

System update is one of the important factors affecting CDSS adoption. The continuous development of the CDSS is necessary in order to avoid any errors and keep abreast of any developments or new knowledge. The refinement of knowledge that is housed in databases, such as updating the medications information in terms of appropriate dosage amounts and side effects, must be done and takes into account the patient's condition and age. The results of the interviews indicate that the maintenance and continuous updating of the system contributes greatly to the successful implementation and sustainability of the CDSS.

> *"Updating the system is very important because the fields of medical knowledge are constantly developing, so the system needs constant updating."* (GP13)

> *"Updating the system is very important. I mean sometimes with time the user will discover some issues that need to be fixed or updated. For example, regarding the current system of HIS, I see things that need development."* (GP23)

CDSS is considered to be a knowledge-based system and it is subsequently necessary to update the system and its database to offer high-quality healthcare to patients, founded on evidence-based medicine (Breighner & Kashani, 2017). With continuous improvement and updating of CDSS, both its own efficiency and the experience of healthcare professionals will also make progress (Meulendijk et al., 2016). Koskela et al. (2016) noted the importance of continuous updating of data in the CDSS, which will make the system easier to use and give more effective results in relation to medical decision-making.

### 4.13 Professional skills and knowledge

When adopting the CDSS, it is important to highlight the extent to which the adoption of the system affects the doctor's professional skills and knowledge. The CDSS contributes to overall healthcare services by providing medical and diagnostic information and advice to help make decisions.

*GP1 'I'm afraid that the technology is completely relied upon, causing a lack of good communication with the patient. In addition, there is something called the art of treatment, which depends on the doctor's view of the patient, which I fear will decrease with the days due to the use of technology, which may affect the skills of the doctor.'*

It is important that the CDSS is designed in a way that recognizes the expertise and capabilities of physicians as it is designed to support and assist the physician's decision, not replace a physician and is built on a rigorous scientific foundation (Shortliffe & Sepúlveda, 2018). The CDSS can affect the skills and knowledge of the end user when relying excessively on the accuracy of the system (Sutton et al., 2020).

### 4.14 Computer efficiency

The presence of high-efficiency hardware devices along with medical software is considered as leading to the better performance of the medical task. One GP in this study believed that having high-quality computers is important and that any defect in them will simply make redundant any medical task in the CDSS.

> *"The presence of modern and high-quality computers contributes to a better performance of medical work. Also, the better the computer the easier is the work, but sometimes it annoys me when I am using the keyboard and sometimes happened a slow or hanging response in the computer."* (GP5)

Zaman et al. (2018) stated in their study that one of the most important obstacles facing the adoption of the HIS in Makkah is the lack of computers and technical expertise. Furthermore, regarding the importance of computer quality, study (Rahi, Khan, and Alghizzawi, 2020) indicated that the effectiveness of computer software can lead to patients accepting telemedicine systems. High-quality computers are among the main assets in hospitals that help to provide better quality services. According to Ammenwerth, Iller, and Mahler (2006), when IT devices are updated, performance is improved, thus increasing the compatibility and fit between task and technology. The authors argued that the presence of quality technology devices, makes that technology more efficient and productive.

### 4.15 Reliability of information

One of the critical factors that influences CDSS adoption is the reliability of the information and recommendations that the system delivers. Any defect in the system or doubt about the data stored in the system will greatly affect adoption of the system and lead to doctors working poorly.

> *"It is important that the information stored in the program is correct and accurate, as sometimes the diagnostic parameters are incomplete or need updating."* (GP11)

> *"There may be some errors in the information provided or even in the data entry, so the calculation of drug doses must be more accurate, for example, providing a calculator specialized in calculating drug doses in the system in order to reduce errors."* (GP22)

Despite the importance of its use and benefits, CDSS is not without risk. When developing CDSS, it is vital to ensure the application is error-free because it is a technology that provides advice and recommendations for patient care. An important factor in CDSS effectiveness could be related to amount of data the system can store.

### 5. Barriers to the adoption of CDSS

This paper aimed to investigate in depth the factors that hinder the adoption of the CDSS by GPs working in primary health center clinics. Views were collected through semi-structured interviews and based on that fifteen factors affecting CDSS adoption were explored that must be taken into consideration by researchers, developers and system providers. The factors are: Time risk, quality of the system used, slow Internet speed, user interface, lack of training, high costs, patient satisfaction, multiple systems used, technical support, computer skills, lack of flexibility, system update, professional skills and knowledge, computer efficiency and quality and reliability of information. The paper provided a great contribution to the current body of HIS adoption research, by exploring what is happening in developing countries (e.g. Saudi Arabia) through in-depth interviews of CDSS adoption. Figure 1 summarizes all the obstacles affecting CDSS adoption.

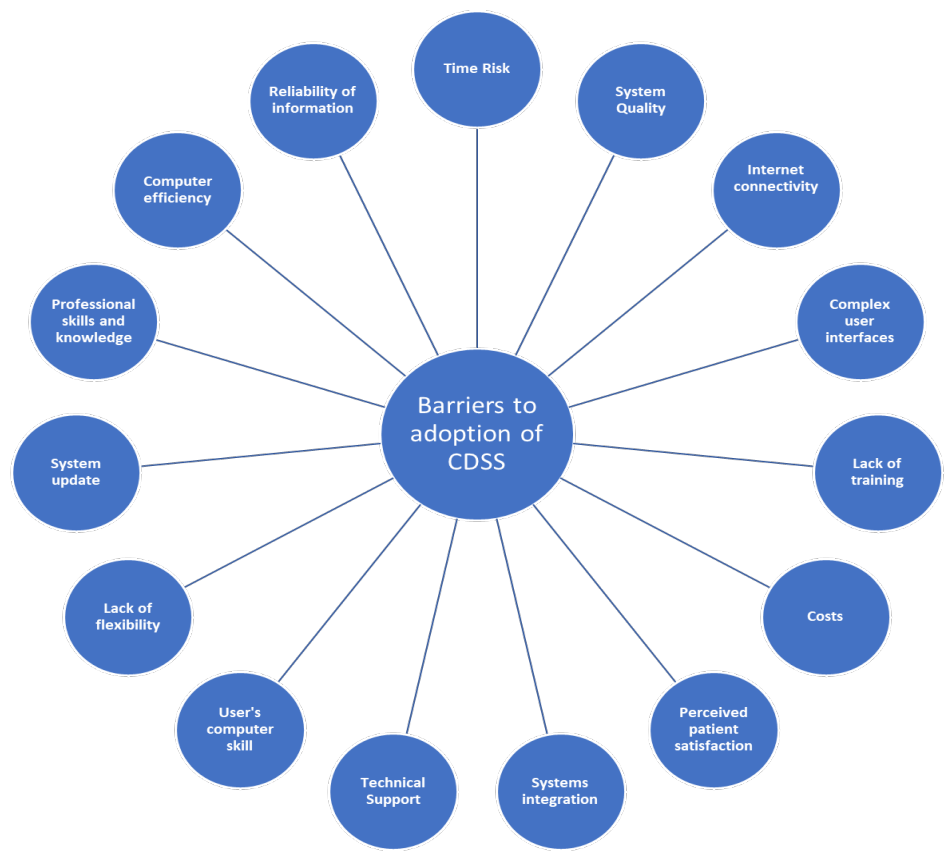

Figure 1: Barriers to the adoption of CDSS

6. Conclusion

This study has contributed to a better understanding of how to enhance the CDSS adoption. This understanding can be applied to other advanced HIS adoption in developing countries healthcare context for identifying challenges. The study benefits HIS researchers and practitioners in the societies generally, especially developing countries to avoid the problems that could happen with the implementation of the CDSS. The paper provided an opportunity for quantitative research further to test and discover new factors in order to study them in more precisely. In addition, the obstacles identified in this study would enable system designers and developers to avoid the problems that may have further impact to compromise the success of CDSS implementation.

This research set up a basis for designing a fully functional CDSS artefact (Miah, Blake and Kerr, 2020) using design research methodology (Miah and Gammack, 2014; Miah, Vu, and Gammack, 2019) addressing different organisational settings (Miah and Yeoh, 2018; Horgan, Ahsan and Miah, 2016) for ensuring implementation success in public healthcare domain. It is anticipated that the design artefact will benefit the healthcare researchers and practitioners in the developing countries where societies are still adopting technologies for effective heath information support.

**References**

Abdekhoda, M., Ahmadi, M., Gohari, M., & Noruzi, A. (2015). The effects of organizational contextual factors on physicians' attitude toward adoption of Electronic Medical Records. *Journal of biomedical informatics, 53*, 174-179

Ajuwon, G. A. (2015). Internet accessibility and use of online health information resources by doctors in training healthcare institutions in Nigeria.

Al Asmri, M., Almalki, M. J., Fitzgerald, G., & Clark, M. (2020). The public health care system and primary care services in Saudi Arabia: a system in transition. *Eastern Mediterranean Health Journal, 26*(4), 468-476. doi:10.26719/emhj.19.049

Al Aswad, A. (2015). Issues concerning the adoption and usage of electronic medical records in ministry of health hospitals in saudi arabia. In: ProQuest Dissertations Publishing.

Alharbi, F., Atkins, A., & Stanier, C. (2015). *Strategic framework for cloud computing decision-making in healthcare sector in Saudi Arabia.* Paper presented at the The seventh international conference on ehealth, telemedicine, and social medicine.

Aljarullah, A., Crowder, R., & Wills, G. (2017). *A framework for the adoption of EHRs by primary healthcare physicians in the kingdom of Saudi Arabia.* Paper presented at International Conference on the Information Society (i-Society), 2017

Alkraiji, A., Jackson, T., & Murray, I. (2016). Factors impacting the adoption decision of health data standards in tertiary healthcare organisations in Saudi Arabia. *Journal of Enterprise Information Management, 29*(5), 650-676.

Alqahtani, S. S., Alshahri, S., Almaleh, A. I., & Nadeem, F. (2016). The Implementation of Clinical Decision Support System: A Case Study in Saudi Arabia. *I.J. Information Technology and Computer Science*, 23-30.

Alshami, M., Almutairi, S., & Househ, M. (2014). The implementation experience of an electronic referral system in saudi arabia: a case study. *Studies In Health Technology And Informatics, 202*, 138-141.

Alsulame, K., Khalifa, M., & Househ, M. (2016). E-Health status in Saudi Arabia: A review of current literature. *Health Policy and Technology, 5*(2), 204-210

Alzaied, T. A. M., & Alshammari, A. (2016). An Evaluation of Primary Healthcare Centers (PHC) Services: The Views of Users. *Health Science Journal, 10*(2), 1.

Amann, J., Blasimme, A., Vayena, E., Frey, D., & Madai, V. I. (2020). Explainability for artificial intelligence in healthcare: a multidisciplinary perspective. *BMC Medical Informatics And Decision Making, 20*(1), 1-9.

Ammenwerth, E., Iller, C., & Mahler, C. (2006). IT-adoption and the interaction of task, technology and individuals: a fit framework and a case study. *BMC Medical Informatics And Decision Making, 6*, 3.

Anderson, C. (2010). Presenting and Evaluating Qualitative Research. *American journal of pharmaceutical education, 74*(8), 1-7.

Arts, D. L., Medlock, S. K., van Weert, H. C. P. M., Wyatt, J. C., & Abu-Hanna, A. (2018). Acceptance and barriers pertaining to a general practice decision support system for multiple clinical conditions: A mixed methods evaluation. *PLoS One, 13*(3), 1-16.

Asan, O., Carayon, P., Beasley, J. W., & Montague, E. (2015). Work system factors influencing physicians' screen sharing behaviors in primary care encounters. *International journal of medical informatics, 84*(10), 791-798.

Bouamrane, M.-M., & Mair, F. S. (2013). A study of general practitioners' perspectives on electronic medical records systems in NHSScotland. *BMC Medical Informatics And Decision Making, 13*(1), 58.

Bouamrane, M.-M., & Mair, F. S. (2014). A qualitative evaluation of general practitioners' views on protocol-driven eReferral in Scotland. *BMC Medical Informatics And Decision Making, 14*(1), 30.

Bologva, E. V., Prokusheva, D. I., Krikunov, A. V., Zvartau, N. E., & Kovalchuk, S. V. (2016). Human-computer interaction in electronic medical records: From the perspectives of physicians and data scientists. *Procedia Computer Science, 100*, 915-920.

Braun, V., & Clarke, V. (2006). Using thematic analysis in psychology. *Qualitative research in psychology, 3*(2), 77-101.

Breighner, C. M., & Kashani, K. B. (2017). Impact of e-alert systems on the care of patients with acute kidney injury. *Best Practice & Research Clinical Anaesthesiology, 31*(3), 353-359.

Castillo, R. S., & Kelemen, A. (2013). Considerations for a successful clinical decision support system. *CIN: Computers, Informatics, Nursing, 31*(7), 319-326.

Chatzakis, I., Vassilakis, K., Lionis, C., & Germanakis, I. (2018). Electronic health record with computerized decision support tools for the purposes of a pediatric cardiovascular heart disease screening program in Crete. *Computer methods and programs in biomedicine, 159*, 159-166.

Dalaba, M. A., Akweongo, P., Aborigo, R. A., Saronga, H. P., Williams, J., Blank, A., . . . Loukanova, S. (2015). Cost-Effectiveness of Clinical Decision Support System in Improving Maternal Health Care in Ghana. *PLoS One, 10*(5), 1-12. doi:10.1371/journal.pone.0125920

Dankovicova, Z., Drotar, P., Gazda, J., & Vokorokos, L. (2018, 2018 / 03 / 27 /). *Overview of the handwriting processing for clinical decision support system.*

De Leeuw, J. A., Woltjer, H., & Kool, R. B. (2020). Identification of Factors Influencing the Adoption of Health Information Technology by Nurses Who Are Digitally Lagging: In-Depth Interview Study. *Journal of medical Internet research, 22*(8), e15630.

Deborah, T. (2014). The Road to Health Data Equity. *Harvard International Review, 35*(4), 48.

Devaraj, S., Sharma, S. K., Fausto, D. J., Viernes, S., & Kharrazi, H. (2014). Barriers and facilitators to clinical decision support systems adoption: A systematic review. *Journal of Business Administration Research, 3*(2), 36.

Esmaeilzadeh, P., Sambasivan, M., Kumar, N., & Nezakati, H. (2015). Adoption of clinical decision support systems in a developing country: Antecedents and outcomes of physician's threat to perceived professional autonomy. *International journal of medical informatics, 84*(8), 548-560.

Etikan, I., Musa, S. A., & Alkassim, R. S. (2016). Comparison of convenience sampling and purposive sampling. *American journal of theoretical and applied statistics, 5*(1), 1-4.

Grout, R. W., Cheng, E. R., Carroll, A. E., Bauer, N. S., & Downs, S. M. (2018). A six-year repeated evaluation of computerized clinical decision support system user acceptability. *International journal of Medical informatics, 112*, 74-81.

Gullapalli, V. K., Brungi, R., & Gopichand, G. (2015). Application of perceptron networks in recommending medical diagnosis. *International Journal of Computer Applications, 113*(4).

Hasanain, R. A., Vallmuur, K., & Clark, M. (2015). Electronic medical record systems in Saudi Arabia: knowledge and preferences of healthcare professionals. *Journal of Health Informatics in Developing Countries, 9*(1).

Holanda, A. A., do Carmo e Sá, H. L., Vieira, A. P. G. F., & Catrib, A. M. F. (2012). Use and Satisfaction with Electronic Health Record by Primary Care Physicians in a Health District in Brazil. *Journal of medical systems, 36*(5), 3141-3149.

Holstiege, J., Mathes, T., & Pieper, D. (2015). Effects of computer-aided clinical decision support systems in improving antibiotic prescribing by primary care providers: a systematic review. *Journal of the American Medical Informatics Association, 22*(1), 236-242.

Hossain, A., Quaresma, R., & Rahman, H. (2019). Investigating factors influencing the physicians' adoption of electronic health record (EHR) in healthcare system of Bangladesh: An empirical study. *International Journal of Information Management, 44*, 76-87.

Horgan, I, Ahsan, K., Miah, S.J. (2016). The Importance of Attributional Trust to Corporate Reputation, Journal of Relationship Marketing 15 (3), 109-134

Irfanahemad, A. S., Nandakumar, B. S., Ugargol, A. P., & Radhika, K. (2018). Factors associated with telemedicine use in a tertiary care pediatrics center - A cross-sectional study. *Al Ameen Journal of Medical Sciences, 11*(01), 31-34.

Jia, P., Zhang, L., Chen, J., Zhao, P., & Zhang, M. (2016). The Effects of Clinical Decision Support Systems on Medication Safety: An Overview. *PLoS One, 11*(12), e0167683. doi:10.1371/journal.pone.0167683

Johnson, R., Evans, M., Cramer, H., Bennert, K., Morris, R., Eldridge, S., . . . Feder, G. (2015). Feasibility and impact of a computerised clinical decision support system on investigation and initial management of new onset chest pain: a mixed methods study. *BMC Medical Informatics And Decision Making, 15*, 71.

Jung, S. Y., Hwang, H., Lee, K., Lee, H.-Y., Kim, E., Kim, M., & Cho, I. Y. (2020). Barriers and Facilitators to Implementation of Medication Decision Support Systems in Electronic Medical Records: Mixed Methods Approach Based on Structural Equation Modeling and Qualitative Analysis. *JMIR Med Inform, 8*(7), e18758.

Keine, D., Zelek, M., Walker, J. Q., & Sabbagh, M. N. (2019). Polypharmacy in an Elderly Population: Enhancing Medication Management Through the Use of Clinical Decision Support Software Platforms. *Neurology and Therapy, 8*(1), 79-94. doi:10.1007/s40120-019-0131-6

Khairat, S., Marc, D., Crosby, W., & Al Sanousi, A. (2018). Reasons For Physicians Not Adopting Clinical Decision Support Systems: Critical Analysis. *JMIR Medical Informatics, 6*(2), e24-e24.

Kitsiou, S., Matopoulos, A., Vlachopoulou, M., & Manthou, V. (2009). Integration Issues in the Healthcare Supply Chain. In *Handbook of Research on Information Technology Management and Clinical Data Administration in Healthcare* (pp. 582-597): IGI Global.

Klarenbeek, S. E., Schuurbiers-Siebers, O. C., van den Heuvel, M. M., Prokop, M., & Tummers, M. (2021). Barriers and facilitators for implementation of a computerized clinical decision support system in lung cancer multidisciplinary team meetings—a qualitative assessment. *Biology, 10*(1), 9.

Koskela, T., Sandström, S., Mäkinen, J., & Liira, H. (2016). User perspectives on an electronic decision-support tool performing comprehensive medication reviews - a focus group study with physicians and nurses. *BMC Medical Informatics And Decision Making, 16*(1), 6. doi:10.1186/s12911-016-0245-z

Kosmisky, D. E., Everhart, S. S., & Griffiths, C. L. (2019). Implementation, Evolution and Impact of ICU Telepharmacy Services Across a Health care System. *Hospital pharmacy, 54*(4), 232-240.

Laka, M., Milazzo, A., & Merlin, T. (2021). Factors That Impact the Adoption of Clinical Decision Support Systems (CDSS) for Antibiotic Management. *International Journal of Environmental Research and Public Health, 18*(4), 1901.

Lodh, N., Sil, J., & Bhattacharya, I. (2017). *Graph Based Clinical Decision Support System Using Ontological Framework*, Singapore.

Lourdusamy, R., & Mattam, X. J. (2020). Clinical Decision Support Systems and Predictive Analytics. In V. Jain & J. M. Chatterjee (Eds.), *Machine Learning with Health Care Perspective: Machine Learning and Healthcare* (pp. 317-355). Cham: Springer International Publishing.

Lugtenberg, M., Weenink, J. W., van der Weijden, T., Westert, G. P., & Kool, R. B. (2015). Implementation of multiple-domain covering computerized decision support systems in primary care: a focus group study on perceived barriers. *BMC Medical Informatics And Decision Making, 15*, 82.

Madar, R., Ugon, A., Ivanković, D., & Tsopra, R. (2021). A Web Interface for Antibiotic Prescription Recommendations in Primary Care: User-Centered Design Approach. *Journal of medical Internet research, 23*(6), e25741.

Maguire, M., & Delahunt, B. (2017). Doing a thematic analysis: A practical, step-by-step guide for learning and teaching scholars. *Journal of Higher Education, 9*(3), 3351-3364

McCloud, R. F., Okechukwu, C. A., Sorensen, G., & Viswanath, K. (2016). Beyond access: barriers to internet health information seeking among the urban poor. *Journal of the American Medical Informatics Association, 23*(6), 1053-1059.

McIntosh, M. J., & Morse, J. M. (2015). Situating and constructing diversity in semi-structured interviews. *Global qualitative nursing research, 2*, 1-12.

Merriam, S. B. (2014). *Qualitative Research. [electronic resource] : A Guide to Design and Implementation*: Hoboken : Wiley, 2014.

3rd ed.

Meulendijk, M. C., Spruit, M. R., Willeboordse, F., Numans, M. E., Brinkkemper, S., Knol, W., . . . Askari, M. (2016). Efficiency of Clinical Decision Support Systems Improves with Experience. *Journal of medical systems, 40*(4), 76. doi:10.1007/s10916-015-0423-z

Miah, S.J., Blake, J, Kerr, D. (2020). Meta-design Knowledge for Clinical Decision Support Systems, Australasian Journal of Information Systems, 24, 1-21

Miah, S.J., Gammack, J. (2014). Ensemble artifact design for context sensitive decision support, Australasian Journal of Information Systems 18 (2), 5-20

Miah, S.J., Yeoh, W. (2018). Applying business intelligence initiatives in healthcare and organizational settings. IGI Global, Pennsylvania, USA. https://doi.org/10.4018/978-1-5225-5718-0

Miah, S.J., Vu, HQ., Gammack, JG (2019). A Location Analytics Method for the Utilization of Geo-tagged Photos in Travel Marketing Decision-Making, Journal of Information and Knowledge Management, 18(1), https://doi.org/10.1142/S0219649219500047

Miller, A., Moon, B., Anders, S., Walden, R., Brown, S., & Montella, D. (2015). Integrating computerized clinical decision support systems into clinical work: A meta-synthesis of qualitative research. *International Journal of Medical informatics, 84*(12), 1009-1018.

Nowell, L. S., Norris, J. M., White, D. E., & Moules, N. J. (2017). Thematic Analysis: Striving to Meet the Trustworthiness Criteria. *International Journal of Qualitative Methods, 16*(1),

Pape, L., Schneider, N., Schleef, T., Junius-Walker, U., Haller, H., Brunkhorst, R., . . . Schiffer, M. (2019). The nephrology eHealth-system of the metropolitan region of Hannover for digitalization of care, establishment of decision support systems and analysis of health care quality. *BMC Medical Informatics And Decision Making, 19*(1), 176.

Payne, T. H., Lovis, C., Gutteridge, C., Pagliari, C., Natarajan, S., Yong, C., & Zhao, L. P. (2019). Status of health information exchange: a comparison of six countries. *Journal of Global Health, 9*(2), 1-16.

Rahi, S., Khan, M. M., & Alghizzawi, M. (2020). Factors influencing the adoption of telemedicine health services during COVID-19 pandemic crisis: an integrative research model. *Enterprise Information Systems*, 1-25.

Rahimi, B., Nadri, H., Lotfnezhad Afshar, H., & Timpka, T. (2018). A Systematic Review of the Technology Acceptance Model in Health Informatics. *9*(3), 604-634. doi:10.1055/s-0038-1668091

Rieckert, A., Sommerauer, C., Krumeich, A., & Sönnichsen, A. (2018). Reduction of inappropriate medication in older populations by electronic decision support (the PRIMA-eDS study): a qualitative study of practical implementation in primary care. *BMC Family Practice, 19*(1), 110.

Ritchie, J., Lewis, J., McNaughton Nicholls, C., & Ormston, R. (2014). *Qualitative research practice : a guide for social science students and researchers*: London : SAGE Publications Ltd, 2014.

Second edition.

Shaker, H. E.-S., & Samir, E.-M. (2014). A distributed clinical decision support system architecture. *Journal of King Saud University: Computer and Information Sciences, Vol 26, Iss 1, Pp 69-78 (2014)*(1), 69.

Shortliffe, E. H., & Sepúlveda, M. J. (2018). Clinical decision support in the era of artificial intelligence. *JAMA Network, 320*(21), 2199-2200.

Stapley, E., Sharples, E., Lachman, P., Lakhanpaul, M., Wolpert, M., & Deighton, J. (2018). Factors to consider in the introduction of huddles on clinical wards: perceptions of staff on the SAFE programme. *International Journal for Quality in Health Care, 30*(1), 44-49.

Sutton, R. T., Pincock, D., Baumgart, D. C., Sadowski, D. C., Fedorak, R. N., & Kroeker, K. I. (2020). An overview of clinical decision support systems: benefits, risks, and strategies for success. *NPJ digital medicine, 3*(1), 1-10.

Tabekhan, A. K., Alkhaldi, Y. M., & Alghamdi, A. K. (2018). Patients satisfaction with consultation at primary health care centers in Abha City, Saudi Arabia. *Journal of Family Medicine and Primary Care, 7*(4), 658.

Thurston, J., & Mulberger, P. K. (2015). Electronic Health Record Implementation Training: Keys to Success. *Nursing Informatics Today, 30*(3), 14-16.

Tsopra, R., Sedki, K., Courtine, M., Falcoff, H., De Beco, A., Madar, R., . . . Lamy, J.-B. (2019). Helping GPs to extrapolate guideline recommendations to patients for whom there are no explicit recommendations, through the visualization of drug properties. The example of AntibioHelp® in bacterial diseases. *Journal of the American Medical Informatics Association, 26*(10), 1010-1019.

Turan, A. H., & Palvia, P. C. (2014). Critical information technology issues in Turkish healthcare. *Information and Management, 51*(1), 57-68.

Wijnhoven, F., & Koerkamp, R. K. (2019). *Barriers for Adoption of Analytical CDSS in Healthcare: Insights from Case Stakeholders.* Paper presented at the 5th International Conference on Information and Communication Technologies in Organizations and Society, ICTO 2019: The Impact of Artificial Intelligence on Business and Society.

Wright, A., Hickman, T. T., McEvoy, D., Aaron, S., Ai, A., Andersen, J. M., . . . Bates, D. W. (2016). Analysis of clinical decision support system malfunctions: a case series and survey. *Journal of the American Medical Informatics Association, 23*(6), 1068-1076.

Yin, R. K. (2014). *Case study research : design and methods*: Los Angeles : SAGE, [2014]

Fifth edition.

Zakane, S. A., Gustafsson, L. L., Tomson, G., Loukanova, S., Sie, A., Nasiell, J., & Bastholm-Rahmner, P. (2014). Guidelines for maternal and neonatal "point of care": needs of and attitudes towards a computerized clinical decision support system in rural Burkina Faso. *International Journal of medical informatics, 83*(6), 459-469.

Zaman, S. B., Hossain, N., Ahammed, S., & Ahmed, Z. (2017). Contexts and opportunities of e-health technology in medical care. *Journal of Medical Research and Innovation, 1*(2), AV1-AV4.

Zaman, T. U., Raheem, T. M. A., Alharbi, G. M., Shodri, M. F., Kutbi, A. H., Alotaibi, S. M., & Aldaadi, K. S. (2018). E-health and its Transformation of Healthcare Delivery System in Makkah, Saudi Arabia. *Health Sciences, 7*(5), 76-82.